\documentclass{iopart}

\usepackage{ifthen}
\usepackage{ifpdf}

\usepackage{latexsym}
\usepackage{amssymb}
\usepackage{bm}

\ifpdf
\usepackage{graphicx}
\usepackage{epstopdf}
\else
\usepackage{graphicx}
\usepackage{epsfig}
\fi

\newcommand{\tbox}[1]{\mbox{\tiny #1}}

\newcommand{\be}[1]{\begin{eqnarray}\ifthenelse{#1=-1}{\nonumber}{\ifthenelse{#1=0}{}{\label{e#1}}}}
\newcommand{\ee}{\end{eqnarray}} 

\newcommand{\hide}[1]{}

\begin{document}

\title[Mesoscopic conductance]{The mesoscopic conductance of disordered rings, \\
its random matrix theory, \\ 
and the generalized variable range hopping picture}

\author{
Alexander Stotland$^{1}$, 
Rangga Budoyo$^{2}$, 
Tal Peer$^{1}$,
Tsampikos Kottos$^{2}$,
and Doron Cohen$^{1}$}

\address{
\mbox{$^{1}$Department of Physics, Ben-Gurion University, Beer-Sheva 84005, Israel} \\
\mbox{$^{2}$Department of Physics, Wesleyan University, Middletown, Connecticut 06459, USA} 
}


\begin{abstract}
The calculation of the conductance of disordered rings requires a theory that goes 
beyond the Kubo-Drude formulation. Assuming ``mesoscopic" circumstances the analysis 
of the electro-driven transitions show similarities with a percolation problem in 
energy space. We argue that the texture and the sparsity of the perturbation matrix 
dictate the value of the conductance, and study its dependence on the disorder strength, 
ranging from the ballistic to the Anderson localization regime. 
An improved sparse random matrix model is introduced to captures the essential 
ingredients of the problem, and leads to a generalized variable range hopping picture.
\end{abstract}

\section{Introduction} 

Closed mesoscopic rings provide the ideal paradigm for testing the manifestation of quantum 
mechanical effects in the mesoscopic realm~\cite{rings,IS,G1,kamenev}. First measurements 
of the conductance of closed rings have been reported more than a decade ago~\cite{orsay},  
{while more recently there is a renewed experimental interest motivated by high precision 
measurements of individual rings~\cite{expr1,expr2}.} 
In a typical experiment a collection of mesoscopic rings is driven by 
a time dependent magnetic flux which creates an electro-motive-force (EMF). 
In what follows we assume low frequency DC {\em noisy} driving (${\omega \sim 0}$)
with power spectrum 
\be{0}
\tilde{F}(\omega) \ \ = \ \ \varepsilon^2 \,\, 
\frac{1}{2\omega_c}\exp\left(-\frac{|\omega|}{\omega_c}\right) 
\ \ \equiv \ \ \varepsilon^2 \delta_{\Gamma}(\omega)
\ee
where $\varepsilon$ is the RMS value of the voltage, 
and the cutoff frequency $\omega_c$ is small 
compared with any relevant semiclassical energy scale, 
but larger compared with the mean level spacing~$\Delta$ 
(note\footnote{Hence there is no issue of quantum recurrences
which would arise for a strictly linear or periodic driving. 
From here on we use units such that ${\hbar=1}$.}).
Optionally, if we had assumed an interaction
with a thermal bath, the role of $\omega_c$ 
would have been played by the level broadening or 
by the temperature \cite{kbv}. 
In such setup one expects the rate of energy absorption 
to be given by Joule's law $G\varepsilon^2$,
where the coefficient $G$ is defined as 
the ``conductance"\footnote{The terminology of this paper, 
and in particular our notion of ``conductance" are the same as in the 
theoretical review \cite{kamenev} 
and in the experimental work \cite{orsay}.}.

As in the linear response theory (LRT) analysis,  
we assume that the coherence time 
is much longer compared with the ballistic time, 
but smaller compared with the Heisenberg time ($1/\Delta$).
But our interest is in what we call {\em mesoscopic circumstances}.
Namely, {\em we assume that the environmental induced relaxation
is a slow process, when compared with the EMF-driven transitions}  
[this is in opposite to the LRT limit $\varepsilon\rightarrow0$]. 
Accordingly we have to work within the framework 
of {\em semi-linear response theory} (SLRT) \cite{kbr,bls,slr}\footnote{
The term ``semi-linear response" 
to describe the outcome of the theory of Ref.\cite{kbr} 
has been coined in a subsequent work~\cite{slr} where 
it has been applied to the analysis of the absorption 
of low frequency radiation by metallic grains.}:
This theory (see section~5 below) goes beyond the conventional 
framework of the Kubo formalism.


For diffusive rings the Kubo formalism 
leads to the Drude formula for~$G$.  
A major challenge in past studies 
was to calculate the weak localization 
corrections \cite{G1} to the Drude result, 
taking into account the level statistics  
and the type of occupation \cite{kamenev}.
These corrections are of order $\Delta/\omega_c$ 
and accordingly do not challenge the leading 
order Kubo-Drude result. 
It is just natural to ask
what happens to the Drude result if the disorder 
becomes weak (ballistic case) or strong 
(Anderson localization case). 
In the latter case there are two conflicting 
results for the noise $\omega_c$ dependence of~$G$, 
both following Mott's work \cite{mott}.
The question is whether to regard the noise 
as ``low frequency driving" or as ``temperature". 
On the one hand, on the basis of the Kubo formula, 
one expects a crossover from $G \sim \exp(-L/\ell_{\infty})$ 
[where $\ell_{\infty}$ is the localization length],  
to the noise dependent result $G \sim \omega_c^2 |\log(\omega_c)|^{d{+}1}$, 
where ${d{=}1}$ for quasi one-dimensional (1D) ring.
On the other hand, on the basis of the 
variable range hoping (VRH) {picture},   
one expects $G \sim \exp(-(\omega_0/\omega_c)^{1/{d{+}1}})$, 
where $\omega_0$ is a constant. 
Eventually \cite{kbv} it has been realized that 
both the ballistic, the diffusive and the strong 
localization regimes should be handled on equal footing 
using SLRT. The Kubo theory applies in the LRT limit $\varepsilon\rightarrow0$,
while in mesoscopic circumstances SLRT leads  
to a resistor network \cite{miller} ``hopping" picture 
in energy space that generalizes the real space hopping 
picture of Refs.\cite{ambeg,pollak}.

\section{Outline} 

In this Communication we analyze, within the framework of SLRT, 
the dependence of the mesoscopic conductance of a quasi-1D ring 
on the strength of the disorder. We explain that for both weak 
and strong disorder the non-ergodicity of the quantum eigenstates 
implies having {\em texture} and {\em sparsity} in the perturbation matrix.
{Such features imply that the rate of energy absorption 
is suppressed enormously because the system cannot execute 
{\em connected sequences of transitions}. The implied deviations from 
the Kubo-Drude result are demonstrated numerically in Fig.~1.} 

We introduce a novel random matrix theory (RMT) model,  
with either {\em log-box} or {\em log-normal} distributed elements, 
that captures the essential features of the perturbation matrix. 
A generalized resistor network analysis for the EMF-driven 
transitions in energy space leads to a generalized VRH picture 
in the strong disorder limit, while handling on equal footing 
the opposing limit of very weak disorder.  

{In the first part of this communication (sections~1-5) 
we provide the essential details on the model and 
on the LRT/SLRT calculation, leading to the numerical results 
in section~6. The second part of this communication leads 
to the RMT modeling and to the implied generalized VRH picture.}

\section{Modeling}

We consider disorder quasi-1D ring geometry. 
The amount of disorder is traditionally 
characterized by the mean free path $\ell$. 
The semiclassical theory of the conductance 
leads to the Drude formula\footnote{
{Optionally if the ring is characterized 
by its transmission $g$ then $\ell/L$ is replaced by $g/(1{-}g)$. 
See Ref.\cite{kbr,bls} for details. 
As could be expected the result is in agreement with 
the Landauer theory~\cite{stone} provided ${g\ll1}$. 
Indeed the Landauer formula can be obtained 
from Kubo formula, using a semiclassical evaluation 
of velocity-velocity correlation function, 
whenever the contribution of trajectories 
with non-zero winding number can be neglected.}}
\be{4}
G_{\tbox{Drude}} = \frac{e^2}{2\pi\hbar} \mathcal{M} \ \frac{\ell}{L}
\ee
where $L$ is the length of the ring, and $\mathcal{M}$ is the number of open modes 
(proportional to its cross section). 
For the numerical calculations we have used the 
Anderson tight binding model, where the lattice is of size ${L\times M}$ 
with $M \ll L$. The longitudinal hopping amplitude per unit time 
is ${c_{\parallel}=1}$, while in the transverse direction it 
is numerically convenient to have ${c_{\perp}<1}$,  
so as to have in the middle of the band a finite 
energy window with ${\mathcal{M}=M}$ open modes. 

The random on-site potential in the Anderson 
tight binding model is given by a box distribution of width~$W$. 
The density of states at the Fermi energy, 
and the mean free path in the Born approximation,  
are written as 
\be{0}
\varrho_{\tbox{F}} 
\equiv 
\mathcal{M} \frac{L}{\pi\hbar v_{\tbox{F}}} 
\equiv \frac{1}{\Delta}
;\quad\quad\quad 
{\ell\sim \left(\frac{v_{\tbox{F}}}{W}\right)^2}
\ee
The implied definition of $v_{\tbox{F}}$ leads 
to its identification as the Fermi velocity
in the absence of disorder (disregarding a prefactor of order unity).
The Anderson localization length for ${L=\infty}$ sample 
would be ${\ell_{\infty}= \mathcal{M}\ell}$. 
Accordingly, for finite sample, depending on the strength 
of the disorder, one distinguishes between the ballistic regime (${L\ll\ell}$) 
the diffusive regime (${\ell \ll L \ll \ell_{\infty}}$) 
and the Anderson strong localization regime (${L \gg \ell_{\infty}}$).

\section{The LRT calculation}

The Fluctuation-Dissipation version of the Kubo formula expresses the conductance as 
an integral over the velocity-velocity correlation function: 
\be{13}
G_{\tbox{LRT}} \ \ = \ \ \varrho_{\tbox{F}}  \left(\frac{e}{L}\right)^2 \times \frac{1}{2}\int_{-\infty}^{\infty} \langle v(t) v(0) \rangle dt
\ee
where $v$ is the velocity in the longitudinal direction.
The Drude formula Eq.(\ref{e4}) is based  
on the simplest classical approximation:
\be{14}
\langle v(t) v(0) \rangle  \ \ \approx \ \  
v_{\tbox{F}}^2 \exp\left[-2\left(\frac{v_{\tbox{F}}}{\ell}\right)|t| \right]
\ee
Our objective is to find the conductance 
of the closed ring in circumstances such 
that the motion inside the ring is coherent 
(quantum interferences are not ignored).
The calculation involves the quantum version 
of $\langle v(t) v(0) \rangle$ which can be obtained 
as the Fourier transform of the spectral function
\be{27}
\tilde{C}(\omega) \ = \   
\frac{1}{N} \sum_{nm} |v_{nm}|^2 \,2\pi\delta_{\Gamma}(\omega-(E_m{-}E_n)) 
\ee 
where $N$ is the size of the energy window of interest.
This spectral function can be re-interpreted 
as describing the band profile of the perturbation matrix $\{v_{nm}\}$. 
{In particular the calculation of 
${\tilde{C}(0) \equiv 2\pi\varrho_{\tbox{F}}\langle\langle |v_{nm}|^2 \rangle\rangle_{\tbox{LRT}}}$
involves a simple {\em algebraic} average over the near diagonal matrix elements 
at the energy range of interest.} Using this notation 
the formula for the Kubo conductance takes the form 
\be{26}
G_{\tbox{LRT}} \ \ = \ \  
\pi
\left(\frac{e}{L}\right)^2  
\varrho_{\tbox{F}}^2 \,\langle\langle |v_{nm}|^2 \rangle\rangle_{\tbox{LRT}}
\ee
The $\mathcal{O}(\Delta/\omega_c)$ weak localization corrections 
to the Drude formula Eq.(\ref{e4}) are determined 
by the interplay of the broadened delta function in Eq.(\ref{e27}) 
with the level statistics. See Ref.\cite{kamenev}. Equivalently we may say 
that the {\em algebraic} average ${\langle\langle \cdots\rangle\rangle_{\tbox{LRT}}}$
has some weak sensitivity to the off-diagonal range of the averaging.

\section{The SLRT calculation}

As in the standard derivation of the Kubo formula, also within the framework of SLRT, the 
leading mechanism for absorption is assumed to be Fermi Golden Rule (FGR) transitions. These 
are proportional to the squared matrix elements $|v_{nm}|^2$ of the velocity operator. Still, 
the theory of \cite{kbr} does not lead to the Kubo formula. This is because the rate of 
absorption depends crucially on the possibility to make {\em connected} sequences of transitions.
It is implied that both the texture and the sparsity of the $|v_{nm}|^2$ matrix play a major 
role in the calculation of~$G$.
SLRT leads to a formula for $G$ that can be cast into the form of Eq.(\ref{e26}), provided 
the definition of ${\langle\langle...\rangle\rangle}$ is modified. 
Following \cite{slr,bls} we regard the energy 
levels as the nodes of a resistor network. We define 
\be{11}
\mathsf{g}_{nm}  = 
2\varrho_{\tbox{F}}^{-3} \ 
\frac{|v_{nm}|^2}{(E_n{-}E_m)^2} \   
\delta_{\Gamma}(E_m{-}E_n)
\ee
Then it is argued that ${\langle\langle|v_{nm}|^2\rangle\rangle_{\tbox{SLRT}}}$ 
is the inverse resistivity of the network. 
It is a simple exercise to verify  
that if all the matrix elements are the same, 
say  $|v_{nm}|^2 = \sigma^2$, 
then $\langle\langle|v_{nm}|^2\rangle\rangle_{\tbox{SLRT}} = \sigma^2$
too. But if the matrix has texture or sparsity 
then $\langle\langle|v_{nm}|^2\rangle\rangle_{\tbox{SLRT}} \ll \langle\langle|v_{nm}|^2\rangle\rangle_{\tbox{LRT}}$.

\section{Numerical results}

It is natural to define the scaled conductance of the ring as follows: 
\be{28}
\tilde{G} \ = \ \frac{G}{(e^2/2\pi\hbar) \mathcal{M}} 
\ = \ 2  \mathcal{M} \times \frac{1}{v_{\tbox{F}}^2} \langle\langle|v_{nm}|^2\rangle\rangle 
\ee
This would be the average transmission per channel, if we had considered open (Landauer) 
geometry. {But for a closed (ring) geometry $\tilde{G}$ is determined 
by the appropriate ``averaging" procedure ${\langle\langle...\rangle\rangle_{\tbox{LRT / SLRT}}}$. 
In the SLRT case the ``averaging" is in fact a resistor network calculation. 
If all the near diagonal elements are comparable in size then SLRT 
will give essentially the same result as LRT. 
More generally ${\langle\langle|v_{nm}|^2\rangle\rangle_{\tbox{SLRT}}}$   
is typically bounded from above by the {\em algebraic} average, and bounded 
from below by the {\em harmonic} average. 
The latter is defined as $\langle\langle X \rangle\rangle_{\tbox{h}} = [\langle1/X\rangle]^{-1}$, 
and reflects ``addition of resistors in series". If the distribution is not too 
stretched then the {\em median}, or the {\em geometric} average, 
or the {\em mixed} average of Ref.\cite{kbr} might provide a good approximation. 
But in general a proper resistor network calculation is required.} 
The resistor network calculation is sensitive 
to the texture and to the sparsity of the perturbation matrix. 
By {\em texture} we mean that the gray-level image of the $v_{nm}$ matrix  
appears to be scarred by structures, rather than being homogeneous. 
Looking on the images of the $v_{nm}$ matrices for various values of~$W$,  
one realizes that both texture and sparsity emerge in the ballistic case, 
while in the strong localization case one observes only sparsity. 
We further expand on the quantitative characterization below.

In Fig.1 we plot the Drude conductance $\tilde{G}_{\tbox{Drude}}$ of Eq.(\ref{e4}), 
and the Kubo conductance $\tilde{G}_{\tbox{LRT}}$, 
together with the mesoscopic conductance $\tilde{G}_{\tbox{SLRT}}$ versus~$W$. 
{We see that outside of the diffusive regime, 
for both weak and strong disorder, the SLRT result is extremely 
small compared with the LRT expectation. This generic behavior 
is related to the sparsity and to the texture of the perturbation matrix, 
which is implied by the statistical properties of the eigenstates.}
The statistical analysis is carried out in Fig.~2, 
while the RMT perspective is tested in Fig.~3.  
The content of Figs.2-3 is further discussed in the following sections.

In order to determine numerically whether the {\em texture} is of any importance 
we simply permute randomly the elements of the $v_{nm}$ matrix along the diagonals, 
and re-calculate $\tilde{G}$ (see the ``untextured" data points in Fig.~3). 
Obviously, by definition $\tilde{G}_{\tbox{LRT}}$ is not affected 
by this numerical maneuver. But it turns out that $\tilde{G}_{\tbox{SLRT}}$ 
is somewhat affected by this procedure in the ballistic regime,  
but still the qualitative results come out the same.   
Accordingly we deduce that the main issue is the {\em sparsity},  
and concentrate below on the RMT modeling of this feature.

In the remainder of this paper, we pave an analytical 
approach to the calculation of the conductance, which will allow us 
to shed some light on these numerical findings. 
First we discuss the familiar diffusive regime where both LRT and SLRT 
should be in agreement with the Drude approximation 
(the latter should become a good approximation for a big sample). 
Then we discuss the departure of SLRT from LRT outside of the diffusive regime, 
which reflects the sparsity of the $v_{nm}$ matrix
due to the non-ergodicity of the eigenfunctions.

\section{The Random Wave conjecture}

In the diffusive regime Mott has argued 
that the eigenstates of the Hamiltonian matrix 
are ergodic in position space, 
and look like random waves. 
Using this assumption 
one can reconstruct the Drude result.
Following Mott\footnote{
The original argument by Mott is somewhat vague. 
We thank Holger Schanz for helpful communication
concerning a crucial step in the derivation.}
we assume that~$\ell$ 
is the correlation scale of 
any typical eigenfunction $\Psi(x,y)$.
The basic assumption of Mott is
that the eigenstates are locally similar
to free waves. The total volume $L^d$
is divided into domains of size $\ell^d$.
Hence we have $(L/\ell)^d$ such domains.
Given a domain, the condition to have
non-vanishing overlap upon integration 
is ${|\vec{q}_n-\vec{q}_m|\ell < 2\pi}$, 
where $\vec{q}$ is the local wavenumber
within this domaim. The probability
that $\vec{q}_n$ would coincide
with $\vec{q}_m$ is ${1/(k_E\ell)^{d{-}1}}$. 
The contributions of the non-zero
overlaps add with random signs hence
\be{0}
|v_{nm}| =
\left[ \frac{1}{(k_E\ell)^{d{-}1}} \times \left(\frac{L}{\ell}\right)^d \right]^{1/2}
\times (\overline{\Psi^2}\ell^d)v_{\tbox{F}}
\ee
where assuming ergodicity $\overline{\Psi^2} \approx 1/L^d$. 
From here we get ${\tilde{G} \sim \ell/L}$ leading to the Drude result. 
We discuss the limited validity of this result in the next section.

\section{The non ergodicity issue}

It is clear that Mott's derivation of the Drude formula on the basis of LRT 
and the Random Wave conjecture becomes non-applicable if the eigenfunctions are non-ergodic. 
This is indeed the case for both weak and strong disorder: 
a typical eigenfunction does not fill the whole 
accessible phase space. In the ballistic regime a typical eigenfunction is not ergodic over 
the open modes in momentum space, while in the strong localization regime it is not ergodic 
over the ring in real space~\cite{ParticRatio}. Fig.2 demonstrate this point by plotting 
the participation ratio as a function of the disorder strength. Lack of quantum ergodicity 
for either weak or strong disorder implies that the perturbation matrix $v_{nm}$ is very 
textured and/or sparse.  For the following analysis a precise mathematical definition  
of sparsity is required. In the next sections we shall provide such a definition, 
but for this purpose we have to shed some light on the size distribution of the matrix elements.

{\em In the strong disorder regime} the observed ``sparsity" is very simple for understanding: 
Eigenstates that are close in energy are typically distant in real space, and therefore have very 
small overlap. The ``big" matrix elements are contributed by eigenstates that dwell in 
the same region in real space, and hence sparse in energy space.  
What we are going to call in the next sections ``sparsity", 
is merely a reflection of the associated log-box size distribution 
of the matrix elements (see Fig.2b). 
The log-box distribution is deduced by a straightforward extension of the above 
argument. A generic eignefunction in the localized regime has an exponential 
shape ${\psi(r) \sim \exp(-|r-r_0|/\ell_{\infty})}$ which is characterized 
by the localization length $\ell_{\infty}$.  
Consequently a typical matrix element of ${ \{|v_{nm}|^2\} }$ has the magnitude 
\be{0}
X \ \ \sim \ \ \frac{1}{\mathcal{M}^2}v_{\tbox{F}}^2 
\,\exp\left(-\frac{x}{\ell_{\infty}}\right)
\ee
where $x\in[0,L/2]$ has a uniform distribution.      
The prefactor is most easily derived from 
the requirement of having ${\langle X \rangle \approx (\ell/\mathcal{M}L) v_{\tbox{F}}^2 }$ 
in agreement with the semiclassical result. 
The latter is deduced from the Fourier transform of 
the velocity-velocity correlation function Eq.(\ref{e14}).

{
{\em In the weak disorder regime} the explanation of the observed ``sparsity" and textures  
requires some more effort. For the purpose of this Communication we shall be 
satisfied with a qualitative explanation:   
If the disorder~$W$ were zero, 
then the mode index (call it~$n_y$) would become 
a good quantum number. This means that states  
that are close in energy are not coupled 
(because they have different~$n_y$).
Once~$W$ becomes non-zero (but still small) the 
mixing is described by Wigner Lorentzians 
(much the same as in the toy model of Ref.\cite{kbr}). 
Then the ratio between small and large couplings 
is determined by the different degree of mixing 
of close versus far modes. 
Consequently one observes a wide (but not stretched) 
distribution for the $\log(X)$ values (see Fig.2b). 
}

\section{RMT modeling, beyond the Gaussian assumption}

{
It was the idea of Wigner~\cite{wigner} to model the perturbation 
matrix of a {\em complex} system as a (banded) random matrix.
Later it has been conjectured by Bohigas~\cite{bohigas} that similar 
modeling may apply to {\em chaotic} systems. 
For many purposes it is convenient to assume infinite bandwidth 
with elements that are taken from a Gaussian distribution, 
leading to the standard Gaussian Orthogonal or Unitary ensembles (GOE/GUE).  
But there are obviously physical circumstance in which  
it is essential to go beyond the Gaussian assumption, 
and to take into account the implications of having finite 
bandwidth and/or non-Gaussian distribution of elements~\cite{prosen}  
and/or sparsity~\cite{Fyodo} and/or texture.}

{
It should be clear that the default assumption of having 
a Gaussian distribution of in-band matrix elements is 
legitimate on practical grounds as long as the matrix elements 
have {\em comparable size in absolute value}. 
But if the eigenfunctions are non-ergodic this assumption 
becomes problematic, because the elements (in absolute value) 
might have a wide distribution over many decades in log scale.
In such case different type of averages may differ by orders of magnitude.}

{
In the following we regard $\{|v_{nm}|^2\}$ 
as a random matrix of non-negative numbers~$\{X\}$.
In general it might be a banded matrix.  
If the standard Gaussian assumption applies  
then the in-band elements of~$\{X\}$ 
are characterized by the Proter-Thomas distribution.
But we are interested in physical circumstances 
in which many of the in-band elements are vanishingly small. 
We define this feature as ``sparsity". 
In the next section we define $p$ as the fraction 
of elements that are larger than the average.
If we have ${p\ll1}$ then we say that the matrix is ``sparse". 
We further discuss the definition of $p$ in the next section.}

\section{Characterization of sparsity}

For an artificially generated sparse random matrix $\{X\}$ 
{of non-negative elements}, 
one defines~$p$ as the fraction of non-zero elements. 
Such a definition assumes a bimodal distribution. 
But in general realistic circumstances   
we do not have a bimodal distribution.  
Rather for strong disorder we already 
had explained that the distribution 
of the matrix elements $\{|v_{nm}|^2\}$ is log-box. 
Contemplating a bit on this issue one concludes 
that the physically generalized 
definition of the sparsity measure 
is ${p\equiv \mbox{F}(\langle X\rangle})$ 
where $\mbox{F}(X)$ is the probability to find 
a value larger than $X$.
We regard a matrix as sparse if $p \ll 1$. 
Given that $\ln(X)$ is uniformly distributed 
between $\ln(X_0)$ and $\ln(X_1)$ 
we define ${\tilde{p} \equiv (\ln(X_1/X_0))^{-1} }$, 
and find assuming $ X_0 {\ll} X_1$ 
that ${p \approx -\tilde{p}\ln\tilde{p}}$,  
and ${\langle X \rangle \approx \tilde{p} X_1}$.
Hence for log-box distribution ${\langle X \rangle \sim p X_1}$, 
as expected from the standard bimodal case.

In Fig.3 we re-do the calculation of the conductance 
with artificial matrices with the same sparsity, 
i.e. log-box distributed elements with the same~$p$. 
We observe qualitative agreement for strong disorder.   
In the other extreme limit of weak disorder there is no 
agreement, because we have to use a different 
distribution for the matrix elements: It turns out that 
also in the ballistic regime $\log(X)$ has a wide distribution, 
but it is not stretched as in the case of a log-box 
distribution {(see Section~8).
In practice we can describe the $X$~distribution of the matrix 
elements in the ballistic limit as log-normal}\footnote{
{The default fitting of the $\log(X)$ distribution 
to a {\em Gaussian} line shape is merely a practical issue. 
The ``RMT ideology" is to see whether a ``minimum information" 
ensemble of random matrices can be used in order to 
derive reasonable estimates. If we want to further improve 
our estimates in the ballistic regime it is essential 
to take into account the texture and not only the deviation 
from log-normal distribution (see Section~6).}}. 
Once we use the appropriate distribution we get 
a reasonable qualitative agreement. 
We emphasize that in both cases,  
of either weak or strong disorder,  
there is besides the algebraic average~$\langle X \rangle$ 
only {\em one} additional fitting parameter 
that characterizes the distribution 
and hence determines the ``sparsity". 
{We could of course have generated RMT matrices 
using the actual distribution (Fig.~2), but then we would 
merely re-generate the {\em untextured} data points.}

Given a hopping range ${|E_m-E_n| \le \omega}$ we can look for 
the typical matrix element $\overline{X}$ for connected sequences 
of such transitions, which we find by solving the equation 
\be{0}
\left(\frac{\omega}{\Delta}\right) \, \mbox{F}\Big(\overline{X}\Big) \ \ \sim \ \ 1 
\ee 
In particular for strong disorder we get
\be{130}
\overline{X} \ \ \approx \ \   
v_{\tbox{F}}^2 \,\exp\left(-\frac{\Delta_{\ell}}{\omega}\right)
\ee 
where $\Delta_{\ell}=(L/\ell_{\infty})\Delta$ is the local level spacing 
between eigenstates that are localized in the same region. 
The same procedure can be applied also in the ballistic 
regime leading to a simpler variation of (\ref{e130}) 
where the dependence of~$\overline{X}$ on~$\omega$ 
predominantly reflects the band profile: It follows from 
the discussion after Eq.(\ref{e27}) that $v_{nm}$ 
is a banded matrix, with a Lorentzian bandprofile 
whose width ${\sim v_{\tbox{F}}/\ell}$ becomes narrower 
as the disorder is decreased.

\section{Generalized Kubo formula}

The definition of the bandprofile reflects the 
variation of $\langle X \rangle$ 
with $\omega$. In complete analogy we define 
an effective bandprofile that reflects 
the variation of $\overline{X}$ with $\omega$. Namely: 
\be{0}
\tilde{C}_{\tbox{qm-LRT}}(\omega)\ \ \ &\equiv& \ \ 2\pi \varrho_{\tbox{F}} \,\langle X \rangle \\ 
\tilde{C}_{\tbox{qm-SLRT}}(\omega) \ \ &\equiv& \ \ 2\pi \varrho_{\tbox{F}} \,\overline{X}
\ee 
The spectral function of Eq.(\ref{e27}) is a smeared 
version of the ``bare" spectral function: it is obtained 
by a convolution $\tilde{C}_{\tbox{qm-LRT}}(\omega)\star\delta_{\Gamma}(\omega)$.
Consequently we get   
\be{0}
G \ = \ 
\frac{1}{2}
\left(\frac{e}{L}\right)^2 
\varrho_{\tbox{F}} \int \tilde{C}_{\tbox{qm}}(\omega) \delta_{\Gamma}(\omega) d\omega 
\ee
where the appropriate LRT/SLRT spectral function should be used. 
This way of writing allows to obtain an approximation 
for the mesoscopic conductance using a Kubo-like calculation:  
it is just re-writing of the Kubo formula in the LRT case, 
while being a generalized VRH approximation in the SLRT case.

For strong disorder the above generalized VRH approximation gives  
an integral over ${\exp(-|\omega|/\omega_c)\exp(-\Delta_{\ell}/|\omega|)}$, 
{which is a product of two competing factors:  
the first has to do with the noise/temperature 
and the second has to do with the couplings. 
This integral is handled using the usual VRH phenomenology: 
the result is determined by the maximum of its integrand,  
which requires to optimize the range~$\omega$ of the transition}.    
In the weak disorder regime the VRH integral 
is not the same as in the strong disorder case, 
because a log-normal rather than log-box 
distribution is involved. 
We have verified that the generalized VRH integral  
gives a qualitatively reasonable approximation 
to the actual resistor network calculation in both cases. 
In any case one should keep in mind 
the well known reservations that apply to  
such ``mean field" approach \cite{ambeg,pollak}.

\section{Summary}

Within SLRT it is assumed that the transitions 
between levels are given by the Fermi-golden-rule (FGR), 
but a resistor network analogy is used in order 
to calculate the energy absorption rate. 
The calculation generalizes the variable range 
hopping picture and treats on equal footing 
the weak and strong disorder regimes.
The essential physics is captured by RMT, provided 
the perturbation matrix is regarded as a member  
of the appropriate Gaussian / Log-Normal / Log-Box ensemble.

{
The prevailing results in the literature 
regarding the conductance of small closed metallic rings 
(for a review see Ref.~\cite{kamenev})
concern mainly the diffusive regime,  
where in leading order the conductance 
is given by the Drude formula,  
and SLRT does not differ much from LRT.
In the present communication 
multi-mode rings in the non-diffusive regime  
are considered seriously for the first time.
Then it become essential to define the precise 
assumptions regarding the environment and the driving.
It is important to realize that both LRT and SLRT 
assume Markovian FGR transitions. This is a very realistic  
assumption that can be justified rigorously if 
one assumes noisy driving (as in our exposition)
or else it is implied by having 
a noisy environment\footnote{ 
{The possibility to witness dynamical localization 
related corrections~\cite{basko} that go beyond the FGR picture, 
requires strictly coherent microscopic-like circumstances, 
such that the dephasing time is much longer 
compared with the {\em Heisenberg time}, and the 
low frequency driving is required to be strictly periodic 
over those extremely long periods.  
Such conditions are possibly not easy to achieve   
in realistic experimental circumstances  
once multi-mode rings are concerned.}}.
Accordingly it should be clear that LRT and SLRT 
both share the same small parameter as in the FGR picture,  
which is the ratio between the rate of the driven transitions 
and the smallest relevant energy scale that characterizes 
the band-profile (bandwidth/sparsity/texture). 
}

{There is only one assumption that distinguishes 
the SLRT (mesoscopic) circumstances from LRT (Kubo) circumstances.}  
This is related to the implicit role 
of the environmentally induced relaxation process 
in the determination the steady state of the system.
Within SLRT one assumes that the FGR rate of the driven transitions
($w_{\tbox{FGR}} \propto \varepsilon^2\mathsf{g}_{nm}$)   
is larger compared with the relaxation rate ($\gamma_{\tbox{rlx}}$). 
The inelastic relaxation effect can be incorporated 
into the SLRT framework by considering a non-symmetric~$\mathsf{g}_{nm}$ 
as implied (say) by detailed balance considerations.  
If the relaxation is the predominant effect 
(${w_{\tbox{FGR}}<\gamma}$) then we are back in 
the LRT regime \cite{more} 
where the Kubo-Drude result applies \cite{kbr}.

{One can wonder what happens if the FGR assumption of LRT/SLRT 
breaks down. Not much is known \cite{rsp}.}  
Ref.\cite{silva} has attempted to go beyond the FGR approximation 
using the Keldish formalism, and has recovered a Marokovian picture 
that leads to a Kubo-like result for the conductance. 
If the Keldish Markovian picture could be established beyond 
the diffusive regime \cite{silvaPC}, 
it would be possible to extend SLRT into the non-linear regime.


\ack

This research was supported by a grant from the USA-Israel
Binational Science Foundation (BSF).




\vspace*{5mm}

\begin{center}
\includegraphics[clip,width=0.65\hsize]{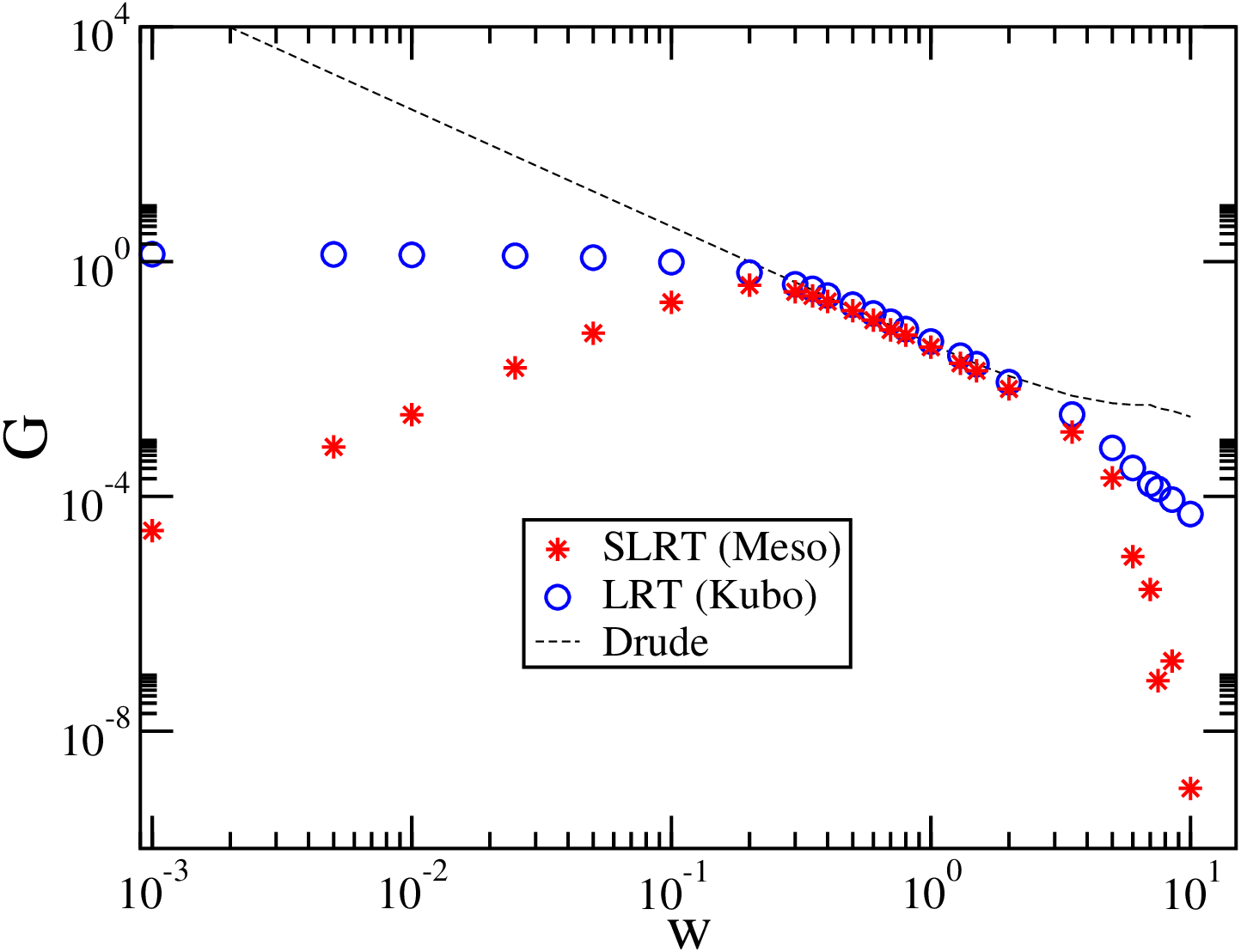}
\end{center}
{\footnotesize {\bf Fig.1:}
Plot of the scaled conductance $\tilde{G}$ versus $W$ 
using either the LRT (Kubo) or the SLRT (mesoscopic) recipe, 
and compared with Drude formula based estimate.
The calculation has been carried out for a tight binding 
Anderson model of size ${500 \times 10}$, 
transverse hopping amplitude ${c{=}0.9}$ 
and low driving frequency $\omega_c/\Delta = 7$.
The SLRT result departs from the LRT result 
for both weak disorder (ballistic regime) 
and strong disorder (strong localization regime).}

\newpage

\begin{center}
\includegraphics[clip,width=0.8\hsize]{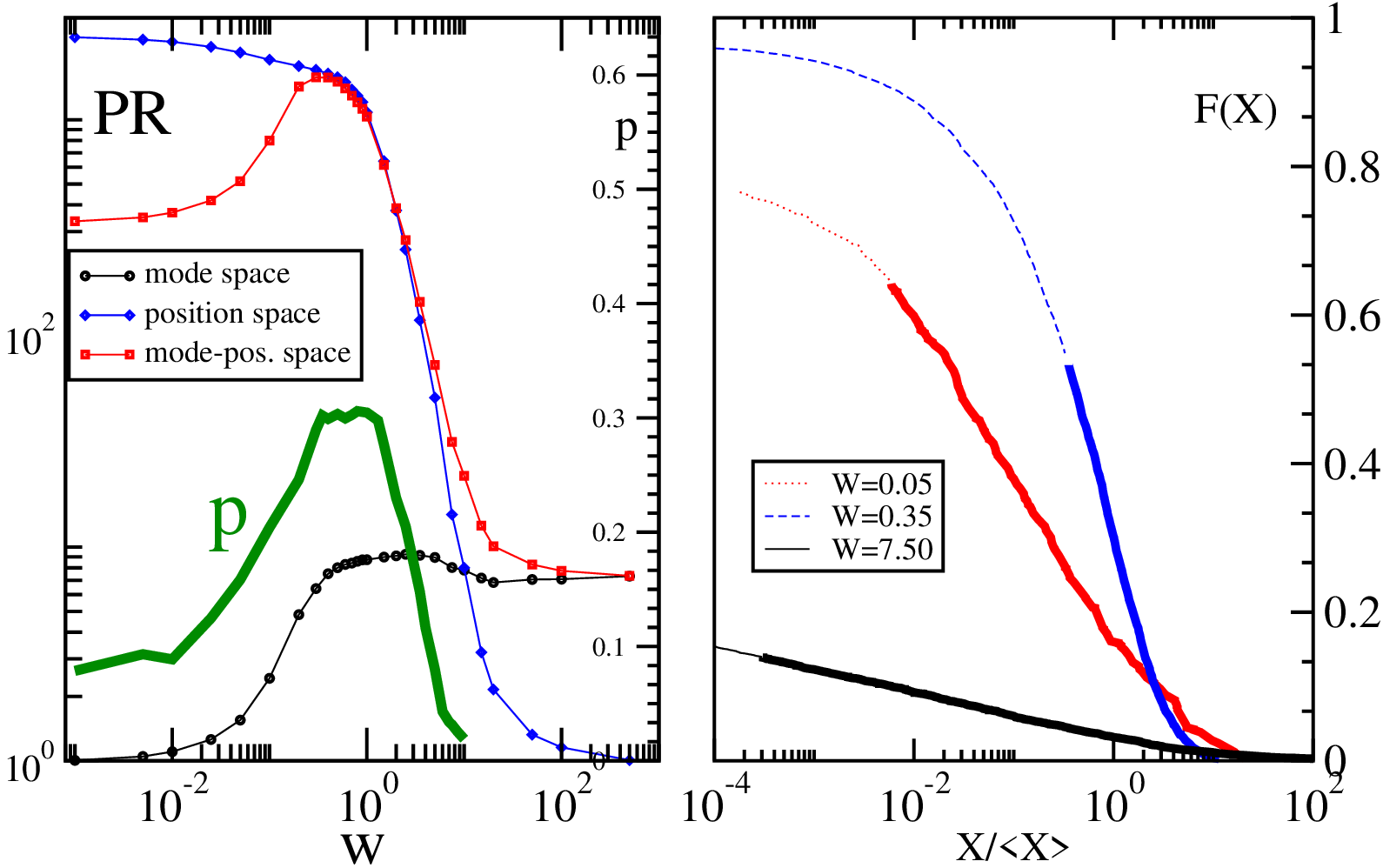}
\end{center}
{\footnotesize {\bf Fig.2:}
The ergodicity of the eigenstates is characterized 
by the participation ratio $\mbox{PR}\equiv[\sum\rho^2]^{-1}$,  
which is calculated (left panel) in various representations: 
in position space $\rho_{r_x,r_y}=|\langle r_x,r_y|\Psi\rangle|^2$, 
in position-mode space $\rho_{r_x,k_y}=|\langle r_x,k_y|\Psi\rangle|^2$,  
and in mode space $\rho_{k_y}=\sum_{r_x}|\langle r_x,k_y|\Psi\rangle|^2$, 
where $k_y=[\pi/(\mathcal{M}{+}1)]\times\mbox{\small integer}$.
The cumulative distribution $\mbox{F}(X)$ of the in-band matrix 
elements (right panel) exhibits a log-box distribution in the 
strong localization regime. 
Points in the interval ${X > \langle\langle X \rangle\rangle_{\tbox{SLRT}}}$,  
corresponding to non-negligible values, are connected by a thicker line. 
The extracted sparsity measure (left panel) is ${p\equiv F(\langle X\rangle)}$.}

\ \\ \ \\

\begin{center}
\includegraphics[clip,width=0.8\hsize]{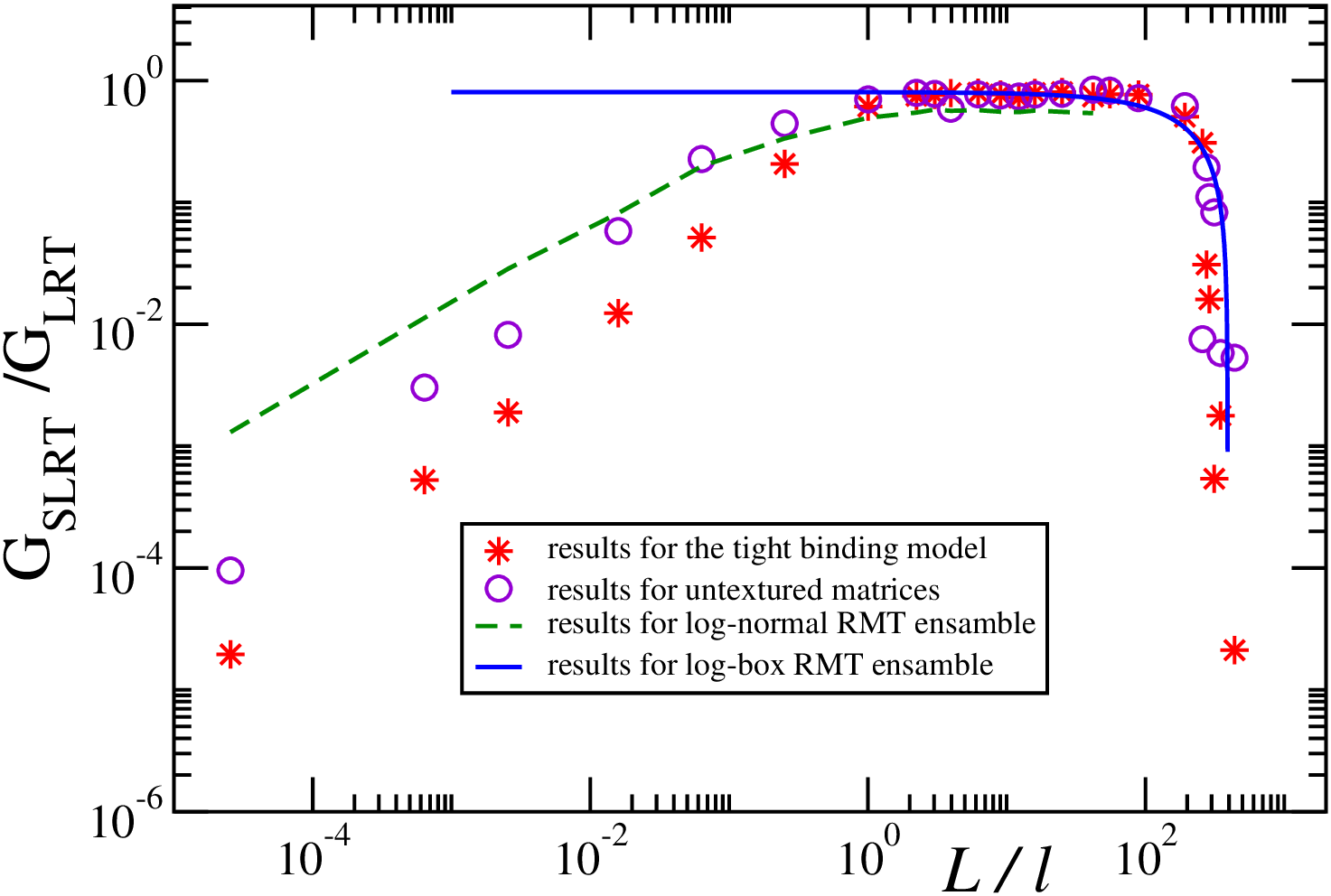}
\end{center}
{\footnotesize {\bf Fig.3:}
The ratio $G_{\tbox{SLRT}}/G_{\tbox{LRT}}$ 
versus the inverse of $\tilde{G}_{\tbox{Drude}}=\ell/L$ 
based on the numerics of Fig.1, 
and compared with artificial RMT modeling 
using ``sparse" matrices formed 
of log-normal or log-box distributed elements. 
We also compare the actual results with 
``untextured" results as explained in the text.  
For weak disorder the agreement is only qualitative 
indicating that the texture becomes important.}

\end{document}